\documentstyle[preprint,prl,aps]{revtex}
\begin{document}
\title{Exactly solvable model of superconducting magnetic alloys}
\author{Valery I. Rupasov\cite{pa}}
\address{Department of Physics, University of Toronto, Toronto,
Ontario, Canada M5S 1A7\\and\\
Landau Institute for Theoretical Physics, Russian Academy of Sciences,
Moscow, Russia}
\date{\today}
\maketitle
\begin{abstract}
A model describing the Anderson impurity in the Bardeen-Cooper-Schriffer
superconductor is proven to exhibit hidden integrability and is
diagonalized exactly by the Bethe ansatz.
\end{abstract}

The basic theoretical models describing magnetic impurities in
nonmagnetic normal metals, such as the s-d (Kondo) model,
the Anderson model, etc., are integrable under {\em two additional
conditions} \cite{MA}: (i) an electron-impurity coupling is assumed
to be energy independent, and (ii) a band electron dispersion
$\varepsilon(k)$ can be linearized around the Fermi level,
$\varepsilon(k)\simeq v_F(k-k_F)$, where $k_F$ and $v_F$ are the
Fermi momentum and velocity, respectively. Only under these conditions,
both an electron-impurity scattering and an effective electron-electron
coupling are described in terms of discontinuous jumps in the Bethe
ansatz wave functions. Therefore a linear dispersion of particles
and a pointlike particle-impurity coupling are considered now
as the {\em necessary} mathematical conditions for integrability
of the ``impurity'' models. Because a carrier dispersion in
superconducting metals cannot be linearized, the ``linear dispersion
approximation'' (LDA) is clear to eliminate superconductivity
from an exact analysis of the behaviour of magnetic alloys \cite{MA}.

It has recently been discerned \cite{RS1} that LDA is not necessary
for an exact diagonalization of the basic impurity models of quantum
optics, describing a system of Bose particles with a nonlinear dispersion
coupled to two-level atoms \cite{RS2}. Making use of some mathematical
analogies between ``magnetic'' and ``optical'' models, it can be shown
\cite{R} that the degenerate Anderson model with a nonlinear band electron
dispersion also exhibits hidden integrability \cite{RS1} and is
diagonalized exactly by the Bethe ansatz. One of the most exciting
applications of the approach developed \cite{RS1,RS2,R} could be an
exact treatment of the superconductivity problem in the presence
of magnetic impurities. Since the pioneering work of Abrikosov and
Gor'kov \cite{AG}, the problem has been the subject of many theoretical
and experimental studies \cite{VIK} but still remains one of the most
intriguing puzzles of modern physics. Therefore an extension
of the Bethe ansatz technique to superconducting magnetic models
accounting for a gap dispersion of charge carriers is of fundamental
physical interest.

In the present letter, we report first results for a model describing
the Anderson impurity placed within the Bardeen-Cooper-Schriffer
(BCS) superconductor. To diagonalize the model Hamiltonian, we
introduce auxiliary particles and show that the multiparticle
scattering process is factorized into two-particle ones, despite
a nonlocal effective particle-impurity coupling. The continuous
multielectron wave functions result from an integral ``dressing''
of the discontinuous Bethe ansatz wave functions of auxiliary particles,
the information about the electron dispersion being contained into
a dressing function. Imposing the periodic boundary conditions on
the multielectron wave functions, we derive the Bethe ansatz equations
(BAE) for the cases of a rare-earth and a transition metal impurity
with infinitely large Coulomb repulsion on the impurity orbital.

To derive the model under consideration, one can start with the
Hamiltonian involving the Hamiltonians of the BCS and Anderson models,
\begin{eqnarray}
H&=&\sum_{\sigma=\uparrow\downarrow}\sum_{{\bf k}}
\varepsilon_{\bf k}a^\dagger_{{\bf k}\sigma}a_{{\bf k}\sigma}
+\sum_{{\bf k}}
\left(\Delta a^\dagger_{{\bf k}\uparrow}a^\dagger_{-{\bf k}\downarrow}+
\Delta^* a_{-{\bf k}\downarrow}a_{{\bf k}\uparrow}\right)\nonumber\\
&&+\sum_{\sigma=\uparrow\downarrow}\sum_{{\bf k}}v_{\bf k}
\left(a^\dagger_{{\bf k}\sigma}d_\sigma+
d^\dagger_\sigma a_{{\bf k}\sigma}\right)
+\epsilon_d\sum_{\sigma=\uparrow\downarrow}d^\dagger_\sigma d_\sigma
+Ud^\dagger_\uparrow d_\uparrow d^\dagger_\downarrow d_\downarrow.
\end{eqnarray}
The Fermi operator $a^\dagger_{{\bf k}\sigma}$ creates a conduction
band electron with the momentum ${\bf k}$, spin $\sigma=\uparrow,\downarrow$
and the energy $\varepsilon_{\bf k}=\epsilon_{\bf k}-\epsilon_F$, where
$\epsilon_{\bf k}$ and $\epsilon_F$ are the kinetic and Fermi
energies. An electron localized on the impurity $d$-orbital with
energy $\epsilon_d$ is described by the Fermi operators $d_\sigma$.
The third term of Eq. (1) represents the hybridization of a band
and a $d$-level electrons with the matrix element $v_{\bf k}$, while
the Coulomb repulsion on the impurity orbital is described
by the last term. Two first terms of Eq. (1) are the standard BCS model,
where the gap $\Delta$ is assumed to result from the Cooper pairing
phenomenon. For our purposes, it is convenient to treat $\Delta$
and $\epsilon_F$ as some free parameters, which have to be found
self-consistently, and to start our analysis with the bare vacuum of
the model defined by $a_{{\bf k}\sigma}|0\rangle=d_\sigma|0\rangle=0$.
In the normal state ($\Delta=0$), the model (1) is reduced to the
Anderson model, which has been diagonalized by Wiegmann \cite{W}
within the LDA.

The BCS part of the model Hamiltonian is diagonalized by the
Bogoliubov-Valatin unitary transform \cite{T} to give
\begin{mathletters}
\begin{equation}
H_{\rm BCS}=\sum_{\sigma}\sum_{{\bf k}}\omega_{\bf k}
c^\dagger_{{\bf k}\sigma}c_{{\bf k}\sigma},
\end{equation}
where $\omega_{\bf k}=-\sqrt{\varepsilon^2_{\bf k}+|\Delta|^2}$,
for $k=|{\bf k}|<k_F$ and
$\omega_{\bf k}=\sqrt{\varepsilon^2_{\bf k}+|\Delta|^2}$ for $k>k_F$.
In terms of new Fermi particles (which we will also call ``electrons''),
the hybridization term takes the form
\begin{equation}
V=v\sum_{\sigma}\sum_{{\bf k}}\left(d^\dagger_\sigma c_{{\bf k}\sigma}+
c^\dagger_{{\bf k}\sigma}d_\sigma\right),
\end{equation}
where we have set $k=k_F$ both in the hybridization matrix element,
$v=v(k_F)$, and in the coefficients of the unitary transform.
In addition, we have omitted in Eq. (2b) the terms
$d^\dagger_\sigma c^\dagger_{{\bf k}\sigma}$ and
$d_\sigma c_{{\bf k}\sigma}$, because they do not conserve the number
of particles and could lead only to insignificant small corrections
in comparison with a contribution of the term (2b).
The bare vacuum of the model is defined now by
$c_{{\bf k}\sigma}|0\rangle=d_\sigma|0\rangle=0$.
To obtain the ground state of the BCS Hamiltonian, one needs thus
to fill all states with $k<k_F$.

Making use of the spherical harmonic representation for band electron
operators \cite{MA}, we arrive at the following one-dimensional
form of the model Hamiltonian:
\end{mathletters}
\begin{mathletters}
\begin{equation}
H=\sum_{\sigma=\uparrow,\downarrow}\int_{-\infty}^{\infty}\frac{dk}{2\pi}
\left\{\omega(k)c^\dagger_\sigma(k)c_\sigma(k)+
v\left[d^\dagger_\sigma c_\sigma(k)+c^\dagger_\sigma(k)d_\sigma\right]
\right\}+\epsilon_d\sum_{\sigma=\uparrow,\downarrow}d^\dagger_\sigma d_\sigma
+Ud^\dagger_\uparrow d_\uparrow d^\dagger_\downarrow d_\downarrow
\end{equation}
where
\begin{equation}
\omega(k)=\left\{
\begin{array}{rl}
-\sqrt{|\Delta|^2+k^2}&,k<0\\
\sqrt{|\Delta|^2+k^2}&,k>0
\end{array}\right.
\end{equation}
and the function $\varepsilon(k)$ is linearized around the Fermi level,
$\varepsilon(k)\simeq v_F(k-k_F)$. We also have set $v_F$ equals to 1
and count the electron momentum $k$ from $k_F$ and the energy
$\epsilon_d$ from $\epsilon_F$. In the normal state, Eqs. (3) are
reduced to the integrable version of the Anderson model. Hereafter,
we restrict ourselves to very large values of $U$, so that the double
occupancy of the $d$-level is unlikely and can be excluded.

The model Hamiltonian (3) describes the behavior of a transition metal
impurity in a superconductor. In the case of rare-earth impurities,
one needs to combine the BCS model and the degenerate Anderson model
\cite{MA}. After analogous manipulations, we then obtain
\end{mathletters}
\begin{equation}
H=\sum_{\alpha}\int_{-\infty}^{\infty}\frac{dk}{2\pi}
\left\{\omega(k)c^\dagger_{\alpha}(k)c_{\alpha}(k)
+v\left[c^\dagger_{\alpha}(k)X_{0\alpha}+
X_{\alpha 0}c_{\alpha}(k)\right]\right\}
+\epsilon_f\sum_{\alpha}X_{\alpha\alpha}.
\end{equation}
Here, the Fermi operator $c^\dagger_{\alpha}(k)$ creates an electron
with the total angular momentum projection $\alpha$ and the momentum
modulus $k$. An impurity is described by the Hubbard operators
$X_{ab}$ with algebra $X_{ab}X_{cd}=\delta_{cb}X_{ad}$, where
the index $a=0,\alpha$ enumerates both the nonmagnetic state ($a=0$)
and the degenerate magnetic states ($\alpha=1,\ldots,n$) of the
impurity with the $f$-level energy $\epsilon_f$. The Coulomb repulsion
on the impurity orbital is assumed to be very large, such the multiple
occupancy is excluded. In the normal state, Eq. (4) is reduced to
the $n$-fold degenerate Anderson model, which has been diagonalized
by Schlottmann \cite{S} within the LDA. In the particular case $n=2$,
the model (4) is equivalent to the nondegenerate model (3) with
infinitely large Coulomb repulsion $U$.

To diagonalize the model Hamiltonian, it is convenient to rewrite
Eq. (4) in terms of electron operators on the ``energy scale'',
$c_\alpha(\omega)=(d\omega(k)/dk)^{-1/2}c_{\alpha}(k)$, with algebra
$\{c_\alpha(\omega),c^\dagger_{\alpha'}(\omega')\}=
2\pi\delta_{\alpha\alpha'}\delta(\omega-\omega')$,
\begin{equation}
H=\sum_{\alpha}\int_{C}\frac{d\omega}{2\pi}
\left\{\omega\,c^\dagger_\alpha(\omega)c_\alpha(\omega)
+V(\omega)\left[c^\dagger_\alpha(\omega)X_{0\alpha}+
X_{\alpha 0}c_\alpha(\omega)\right]\right\}
+\epsilon_f\sum_{\alpha}X_{\alpha\alpha},
\end{equation}
where $V(\omega)=v(d\omega(k)/dk)^{-1/2}$ is energy dependent because
of the nonlinear electron dispersion (3b). The integration contour
$C$ consists of two semi-infinite intervals, $C=C_-\oplus C_+$,
where $C_-=(-\infty,-\Delta)$ and $C_+=(\Delta,\infty)$ correspond
to the lower ($k<0$) and upper ($k>0$) branches of the electron
dispersion.

We look now for one-electron eigenstates of the Hamiltonian (5)
in the form
$$
|\Psi_1\rangle=\sum_{\alpha}A_\alpha\left[{\rm g} X_{\alpha 0}+
\int_{C}\frac{d\epsilon}{2\pi}V(\epsilon)\phi(\epsilon)
c^\dagger_\alpha(\epsilon)\right]|0\rangle,
$$
where $A_\alpha$ is arbitrary. The vacuum state contains no band
electrons and the impurity is in the nonmagnetic state,
$c_\alpha(\epsilon)|0\rangle=X_{a\alpha}|0\rangle=0$.
The Schr\"odinger equation $(H-\omega)|\Psi_1\rangle=0$ then
reads
\begin{mathletters}
\begin{eqnarray}
&&(-i\partial_\tau-\omega)\phi(\tau|\omega)+{\rm g}(\omega)\delta(\tau)=0,\\
&&(\epsilon_f-\omega){\rm g}(\omega)+\int_{-\infty}^{\infty}d\tau
\Gamma(\tau)\phi(\tau|\omega)=0,
\end{eqnarray}
where $\phi(\tau)$ is the Fourier image of the auxiliary wave function
$\phi(\epsilon)$, while the effective particle-impurity coupling
$\Gamma(\tau)$ contains the information about the electron dispersion,
\end{mathletters}
$$
\phi(\tau)=\int_{-\infty}^{\infty}\frac{d\epsilon}{2\pi}
\phi(\epsilon)e^{i\epsilon\tau},\;\;\;
\Gamma(\tau)=\int_{C}\frac{d\epsilon}{2\pi}
V^2(\epsilon)e^{-i\epsilon\tau}.
$$
For $\omega\in C$, one gets ${\rm g}(\omega)=[r(\omega)+i/2]^{-1}$ and
\begin{equation}
\phi(\tau|\omega)=\frac{r(\omega)-(i/2){\rm sgn}(\tau)}{r(\omega)+i/2}
e^{i\omega\tau}.
\end{equation}
The ``rapidity'' $r(\omega)$ is defined by
\begin{equation}
r=\frac{\omega-\epsilon_f}{V^2(\omega)}=
\left(1-\frac{\epsilon_f}{\omega(k)}\right)\frac{k}{v^2},
\end{equation}
where $V^2(\omega)/2$ is the imaginary part of the self-energy
$$
\Sigma(\omega)=\int_{C}\frac{d\epsilon}{2\pi}
\frac{V^2(\epsilon)}{\epsilon-\omega-i0}.
$$
The real part of the self-energy,
$\Sigma'(\omega)={\mbox Re}\,\Sigma(\omega)$, equals to zero
if $\omega$ lies outside the gap, $|\omega|>\Delta$. If $\omega$
lies inside the gap, $\omega\in G=(-\Delta,\Delta)$, the
imaginary part of the self-energy vanishes, and Eqs. (6) admit
also the discrete mode with the eigenenergy $\omega_d$, which is
found as a root of equation $\epsilon_f-\omega=\Sigma'(\omega)$,
where $\Sigma'(\omega)=v^2\omega/2\sqrt{\Delta^2-\omega^2}$ and
$\omega\in G$. The discrete mode describes the electron-impurity
bound state, which is a complete analog of the discrete mode
predicted earlier by John and Wang \cite{JW} for a Bose system.
If $\epsilon_f$ lies far from the gap, $|\epsilon_f|\gg\Delta$,
the bound state energy $\omega_d$ is very close to the conduction
band states, but becomes well separated if $\epsilon_f$ lies inside
and around the gap. Note also that for many superconducting magnetic
alloys $v^2\gg\Delta$, therefore $\omega_d$ lies close to zero
(the Fermi energy of the host metal) for arbitrary position of
the impurity level energy $\epsilon_f$. It should be also
emphasized that the discrete mode is found here as a solution
of the one-particle problem (6) rather than a multiparticle
bound state discussed earlier by Shiba \cite{SH}. Therefore
strong electron-electron correlations could result in
a strong renormalization of the bare discrete mode energy
$\omega_d$.

The auxiliary wave function (7) is discontinuous, but the electron
wave function
$\psi_\alpha(x|\omega)\equiv\langle c_\alpha(x)|\Psi_1\rangle=
\psi(x|\omega)A_\alpha$ (where the operator $c_\alpha(x)$ is defined
as $c_\alpha(x)=\int_{-\infty}^{\infty}(dk/2\pi)c_{\alpha}(k)\exp{(ikx)}$
\cite{MA}) is continuous and results from the integral dressing of
the auxiliary function,
$$
\psi(x|\omega)=\int_{-\infty}^{\infty}d\tau u(x;\tau)\phi(\tau|\omega),
$$
with the dressing function
$$
u(x;\tau)=v\int_{-\infty}^{\infty}\frac{dk}{2\pi}
\left(\frac{d\omega(k)}{dk}\right)^{-1}
\exp{\{i[kx-\omega(k)\tau]\}}.
$$
In the LDA, the dressing function is nothing but the delta-function,
$u(x;\tau)=v\delta(x-\tau)$, and hence the auxiliary wave function
$\phi(\tau|\omega)$ is the same as the electron wave function
$\psi(x|\omega)$.

The idea of auxiliary functions (or ``auxiliary particles'') can
easily be extended to the multielectron case. For instance, the
two-electron wave functions in the energy space are represented
as $\Psi_{\alpha_1\alpha_2}(\epsilon_1,\epsilon_2)=
V(\epsilon_1)V(\epsilon_2)
\Phi_{\alpha_1\alpha_2}(\epsilon_1,\epsilon_2)$ and
$J_{\alpha_1\alpha_2}(\epsilon)=V(\epsilon)
G_{\alpha_1\alpha_2}(\epsilon)$, where the latter describes the state,
in which one of the electrons is localized on the impurity. In
the auxiliary $\tau$-space, the Schr\"odinger equations for the
Fourier images of the auxiliary functions
$\Phi_{\alpha_1\alpha_2}(\tau_1,\tau_2)$ and
$G_{\alpha_1\alpha_2}(\tau)$ are then similar to those in LDA
but with the nonlocal particle-impurity coupling $\Gamma(\tau)$.
It is remarkable that, despite the nonlocal coupling, the two-particle
scattering matrix is still found \cite{RS1,RS2,R} in the well-known
form:
\begin{equation}
{\bf S}=\frac{r(\omega_1)-r(\omega_2)+i{\bf P}}{r(\omega_1)-r(\omega_2)+i}
\end{equation}
where ${\bf P}=\delta_{\alpha_1\alpha'_2}\delta_{\alpha_2\alpha'_1}$
is the permutation operator. The $S$-matrix satisfies obviously the
Yang-Baxter equations. Hence the multiparticle scattering is factorized
into two-particle ones and the auxiliary multiparticle wave functions
have the ordinary Bethe ansatz structure. But due to the nonlinear
dispersion, the multielectron wave functions
\begin{equation}
\Psi_{\alpha_1\ldots\alpha_N}(x_1,\ldots,x_N)=\int_{-\infty}^{\infty}
\Phi_{\alpha_1\ldots\alpha_2}(\tau_1,\ldots,\tau_N)
\prod_{j=1}^{N}u(x_j;\tau_j)d\tau_j
\end{equation}
are continuous functions of the coordinates $x_j$. The factorization
of multielectron scattering and the Bethe ansatz construction for
multielectron wave functions are thus {\em hidden} and manifested
only in the limit of large interelectron separations.

Imposing the periodic boundary conditions on the $N$-electron wave
function (10) on the interval of size $L$, we arrive at the following
Bethe ansatz equations (BAE):
\begin{mathletters}
\begin{eqnarray}
&&\exp{(ik_jL)}e_1(\lambda^{(0)}_j)=
\prod_{a_1=1}^{M_1}e_1(\lambda^{(0)}_j-\lambda^{(1)}_{a_1})\\
&&\prod_{\nu=\pm 1}
\prod_{a_{l+\nu}=1}^{M_{l+\nu}}
e_1(\lambda^{(l)}_{a_l}-\lambda^{(l+\nu)}_{a_{l+\nu}})
=-\prod_{b=1}^{M_l}e_2(\lambda^{(l)}_{a_l}-\lambda^{(l)}_{b})\\
&&E=\sum_{j=1}^{N}\omega_j
\end{eqnarray}
where $E$ is the eigenenergy, $\omega_j\equiv\omega(k_j)$ is the energy
of a ``charge'' excitation of the system with the momentum $k_j$,
and $e_n(x)=(x-in/2)/(x+in/2)$. If $m_\alpha$ is the number of
electrons with the ``color'' $\alpha$, the numbers $M_l$ are defined
by $M_l=\sum_{\alpha=l}^{n-1}m_\alpha$, $M_0$ being the total number
of electrons, $M_0\equiv N$. It is clear that only charge excitations
with rapidities $\lambda^{(0)}_j\equiv r(k_j)$ contain the information
about the nonlinear dispersion, while BAE for the color rapidities
$\{\lambda^{(l)}\}$, $l=1,\ldots,n-1$ coincide with the corresponding
equations in the LDA.

In the particular case $n=2$, Eqs. (11) are reduced to the BAE
of the nondegenerate model (3) with infinitely large $U$,
\end{mathletters}
\begin{mathletters}
\begin{eqnarray}
&&\exp{(ik_jL)}\,\frac{r_j-i/2}{r_j+i/2}=\prod_{\alpha=1}^{M}
\frac{r_j-\lambda_\alpha-i/2}{r_j-\lambda_\alpha+i/2}\\
&&\prod_{j=1}^{N}\frac{\lambda_\alpha-r_j-i/2}{\lambda_\alpha-r_j+i/2}=
-\prod_{\beta=1}^{M}\frac{\lambda_\alpha-\lambda_\beta-i}
{\lambda_\alpha-\lambda_\beta+i},
\end{eqnarray}
where the spin projection of the system is given by $S^z=N/2-M$.
In the normal state, where $\omega=k$ and $r(k)=(k-\epsilon_{f,d})/v^2$,
Eqs. (11) and (12) are the same as the well-known BAE of the
Anderson model.

In conclusion, it should be emphasized that the approach
developed in the present paper for superconducting magnetic
alloys is easily generalized to Fermi systems with nonlinear
spectrum of particles around the Fermi level. For instance,
the BAE derived here are valid also for gapless Fermi systems
\cite{GFS}, where the function $r(k)$ takes the form
\end{mathletters}
\begin{equation}
r(k)=\frac{\omega(k)-\epsilon_f}{v^2}\frac{d\omega(k)}{dk},
\end{equation}
and $\omega(k)$ is the dispersion of particles of the system.

I am grateful to S. John and P. Wiegmann for stimulating
discussions. I would like also to thank the Department
of Physics at the University of Toronto for kind hospitality
and support during the completion of this work.

\end{document}